\documentclass[draft,showpacs,eqsecnum,nofootinbib,aps]{revtex4}

\def\beq{\begin{equation}}
\def\eeq{\end{equation}}
\def\bea{\begin{eqnarray}}
\def\eea{\end{eqnarray}}
\def\nn{\nonumber}
\def\pr{\prime}

\begin{document}
\title{Can wormholes have negative temperatures?}
\author{Soon-Tae Hong}
\email{soonhong@ewha.ac.kr}
\author{Sung-Won Kim}
\email{sungwon@ewha.ac.kr} \affiliation{Department of Science
Education, Ewha Womans University, Seoul 120-750 Korea}
\date{\today}

\begin{abstract}
We study (3+1) Morris-Thorne wormhole to investigate its higher
dimensional embedding structures and thermodynamic properties.  It
is shown that the wormhole is embedded in (5+2) global embedding
Minkowski space. This embedding enables us to construct the
wormhole entropy and wormhole temperature by exploiting Unruh
effects.  We also propose a possibility of negative temperature
originated from exotic matter distribution of the wormhole.
\end{abstract}
 \maketitle

\noindent {\it 1. Introduction.}  Since the cosmic microwave
background was discovered, there have been many ideas and
proposals to figure out how the universe has evolved.  The
standard big bang scenario has led to the inflationary
cosmology~\cite{guth} and nowadays to the M-theory cosmology with
bouncing universes~\cite{seiberg}.  There have been also
considerable discussions on the theoretical existence of wormhole
geometry, since Morris and Thorne (MT) proposed a possibility of
traversable wormhole, through which observers can pass travelling
between two universes as a short cut~\cite{mt88}.  According to
the Einstein field equations, the MT wormhole needs the exotic
matter, which violates the weak energy condition.

On the other hand, it has been discovered the novel aspects that
the thermodynamics of higher dimensional black holes can often be
interpreted in terms of lower dimensional black hole
solutions~\cite{high}.  In fact, a slightly modified solution of
(2+1) dimensional Banados-Teitelboim-Zanelli black
hole~\cite{btz,cal} yields a solution to the string theory,
so-called the black string~\cite{horowitz93}.  Since the thermal
Hawking effects~\cite{hawking} on a curved manifold were studied
as Unruh effects~\cite{unruh} in a higher flat dimensional
space-time, following the global embedding Minkowski space (GEMS)
approach~\cite{kasner} several authors recently have shown that
this approach could yield a unified derivation of temperature for
various curved manifolds in (2+1)
dimensions~\cite{des,btz,cal,hong00prdbtz} and in (3+1)
dimensions~\cite{des,rea,kps00}.  Moreover, the MT wormhole has
been described in terms of its embedding profile surface
geometry~\cite{mt88}.

In this paper we will analyze the geometries of the MT wormhole
manifolds~\cite{mt88} to  construct their higher dimensional flat
embeddings, which will be shown to be related with the embedding
profile surface geometry of the wormhole.  In these GEMS
embeddings, we will investigate the Hawking temperature and
entropy via the Unruh effects to propose a possibility of
``negative temperature" associated with the ``exotic matter."
Recently, the exotic matter was introduced in the
Friedmann-Robertson-Walker model where the bouncing universe could
be initiated by the negative energy density at the big
crunch~\cite{hwang}.  Moreover, a possibility of negative
temperature has been proposed in the de Sitter geometry even
though it was forbidden due to thermodynamic
instability~\cite{minic}.\\

\noindent {\it 2. GEMS geometries of wormholes.}  In order to
study the GEMS structure for wormholes, we start with the static
MT wormhole four-metric of the form~\cite{mt88,kim01} \beq
ds^{2}=e^{2\Phi
(r)}dt^{2}-\left(1-\frac{b(r)}{r}\right)^{-1}dr^{2}
-r^{2}(d\theta^{2}+\sin^{2}\theta d\phi^{2}) \label{staticmetric}
\eeq where the arbitrary smooth functions $\Phi (r)$ and $b(r)$
are the lapse and wormhole shape functions.  Note that, in order
for the wormhole to be maintained, the wormhole function $b(r)$
should be positive and satisfy the flaring-out condition, which
will be explicitly given later in the GEMS structure.  The lapse
function $\Phi (r)$ is finite everywhere.

After some algebra, we obtain the (5+2) GEMS structure
$ds^{2}=\eta_{MN}dx^{M}dx^{N}$ with the flat Minkowski metric \beq
\eta_{MN}={\rm diag}~(+1,-1,-1,-1,-1,-1,+1), \label{ds} \eeq where
the coordinate transformations are given, with two additional
space-like and one time-like dimensions, as follows \bea
x^{0}&=&k_{S}^{-1}e^{\Phi(r)}\sinh{\kappa}t,\nn\\
x^{1}&=&k_{S}^{-1}e^{\Phi(r)}\cosh {\kappa}t,\nn\\
x^{2}&=&r\sin\theta\cos\phi,\nn\\
x^{3}&=&r\sin\theta\sin\phi,\nn\\
x^{4}&=&r\cos\theta,\nn\\
x^{5}&=&\int\frac{{\rm d}r}{\left(1-\frac{b(r)}{r}\right)^{1/2}},\nn\\
x^{6}&=& \int{\rm d}r~\left[1+k_{S}^{-2}(\Phi^{\prime}(r))^{2}
e^{2\Phi(r)}\right]^{1/2}, \label{staticgems} \eea where the
coordinate $x^{5}\in(-\infty,+\infty)$ corresponds to the proper
radial distance measured by static observers~\cite{mt88} and
$k_{S}$ is the surface gravity, which will be discussed later.

For a submanifold on an equatorial slice $(\theta=\pi /2)$ at a
fixed moment of time, the MT wormhole metric (\ref{staticmetric})
is reduced into \beq
ds^{2}=-\left(1-\frac{b(r)}{r}\right)^{-1}dr^{2} -r^{2}d\phi^{2},
\label{staticmetric2} \eeq and its (3+1) GEMS structure is given
by $ds^{2}=-(dx^{2})^{2}-(dx^{3})^{2}-(dx^{5})^{2} +(dx^{6})^{2}$
with \bea
x^{2}&=&r\cos\phi,\nn\\
x^{3}&=&r\sin\phi,\nn\\
x^{5}&=&\int\frac{{\rm d}r}{\left(1-\frac{b(r)}{r}\right)^{1/2}},\nn\\
x^{6}&=&r. \label{staticgems2} \eea Moreover, defining the new
coordinate $z$ as $dz^{2}=(dx^{5})^{2} -(dx^{6})^{2}$, the above
(3+1) GEMS structure is reduced into the (3+0) GEMS one \beq
ds^{2}=-(dx^{2})^{2}-(dx^{3})^{2}-dz^{2}=-dr^{2}-r^{2}d\phi^{2}-dz^{2}
\eeq with $(x^{2},x^{3})$ given by (\ref{staticgems2}) and $z$
defined as \beq z=\int\frac{{\rm
d}r}{\left(\frac{r}{b(r)}-1\right)^{1/2}}, \label{dzdr} \eeq which
describes the embedding profile surface of the MT wormhole
geometry~\cite{mt88}.\\

\noindent {\it 3. Thermodynamics in wormhole GEMS.}  Consider the
thermodynamic properties of the static MT wormhole described by
the GEMS coordinate transformations (\ref{staticgems}).
Introducing the Killing vector $\xi=\partial_{t}$ we evaluate the
surface gravity $k_{S}$ at radius $r$ \beq
k_{S}=\Phi^{\pr}(r)e^{\Phi
(r)}\left(1-\frac{b(r)}{r}\right)^{1/2}. \label{surfaceg} \eeq In
the GEMS structure, the Hawking temperature is attainable through
the relation $T_{H}=a_{7}/2\pi$ with the (5+2) acceleration
$a_{7}$. For the Unruh detectors moving according to constant
$r,~\theta,~\phi$ as in the Schwarzschild black hole
GEMS~\cite{des,kps00}, $a_{7}$ is described by the Rindler-like
motion in the embedded flat space as follows \beq
a_{7}^{-2}=(x^{1})^{2}-(x^{0})^{2} \eeq in the $(x^{0},x^{1})$
plane to yield \beq a_{7}^{-2}=k_{S}^{-2}e^{2\Phi (r)}. \label{a7}
\eeq Substituting the surface gravity $k_{S}$ in (\ref{surfaceg})
into (\ref{a7}), we obtain the Hawking temperature \beq
T_{H}=\frac{a_{7}}{2\pi}=\frac{1}{2\pi}\Phi^{\pr}(r)
\left(1-\frac{b(r)}{r}\right)^{1/2}, \label{hawking} \eeq which is
consistent with the fact that the $a_{7}$ is also attainable from
the relation \beq a_{7}=\frac{k_{S}}{g_{00}^{1/2}}. \eeq Moreover,
the desired wormhole temperature is given by \beq
T_{0}=\frac{k_{S}}{2\pi}=\frac{1}{2\pi}\Phi^{\pr}(r)e^{\Phi (r)}
\left(1-\frac{b(r)}{r}\right)^{1/2}.\label{t0}\eeq  Here one notes
that, even though the event horizon does not exist in the
wormhole, in constructing the above Hawking and wormhole
temperatures we have taken the limit that $r$ approaches $r=b(r)$,
corresponding to the event horizons of the black holes, to yield
the suprema of these wormhole temperatures.  This continuous
limiting procedure makes sense in ensuring the existence of the
Hawking and wormhole temperatures since the temperatures of the
geometrical objects are purely defined in terms of their
geometrical surface gravities~\cite{wald} and the suprema of the
wormhole temperatures are well-defined.

In the orthonormal basis with the metric
$g_{\hat{\mu}\hat{\nu}}={\rm diag}~(+1,-1,-1,-1)$, the wormhole
stress-energy tensor is given by~\cite{mt88}, \beq
T_{\hat{\mu}\hat{\nu}}={\rm diag}(\rho(r), -\tau(r), p(r), p(r)),
\eeq where $\rho$ is the total density of mass-energy, $\tau$ is
the tension per unit area measured in the radial direction, and
$p$ is the pressure measured in lateral direction. Note that, in
an ordinary perfect fluid, $-\tau=p$. In the orthonormal basis,
one can evaluate the non-vanishing Ricci components and the
Einstein curvature: \bea
R_{\hat{0}\hat{0}}&=&\left(\Phi^{\pr\pr}+(\Phi^{\pr})^{2}\right)\left(1-\frac{b}{r}\right)
+\Phi^{\pr}\left(-\frac{3b}{2r^{2}}-\frac{b^{\pr}}{2r}+\frac{2}{r}\right),
\nonumber\\
R_{\hat{1}\hat{1}}&=&-\left(\Phi^{\pr\pr}+(\Phi^{\pr})^{2}\right)\left(1-\frac{b}{r}\right)
+\left(\Phi^{\pr}+\frac{2}{r}\right)\left(-\frac{b}{2r^{2}}+\frac{b^{\pr}}{2r}\right),
\nonumber\\
R_{\hat{2}\hat{2}}&=&R_{\hat{3}\hat{3}}=\frac{1}{r^{2}}\left(\Phi^{\pr\pr}(-r+b)+\frac{b}{2r}
+\frac{b^{\pr}}{2}\right),\nonumber\\
R&=&2\left(\Phi^{\pr\pr}+(\Phi^{\pr})^{2}\right)\left(1-\frac{b}{r}\right)
+\Phi^{\pr}\left(-\frac{3b}{r^{2}}-\frac{b^{\pr}}{r}+\frac{4}{r}\right)-\frac{2b^{\pr}}{r^{2}}.
\label{ricciein} \eea From the Einstein field equations
$G_{\hat{\mu}\hat{\nu}}=R_{\hat{\mu}\hat{\nu}}-\frac{1}{2}
g_{\hat{\mu}\hat{\nu}}R=8\pi G T_{\hat{\mu}\hat{\nu}}$ and
(\ref{ricciein}), one can obtain \bea
b^{\prime}&=&8\pi G r^{2}\rho,\nonumber\\
\Phi^{\prime}&=&\frac{b-8\pi G r^{3}\tau}{2r(r-b)},\nonumber\\
\tau^{\prime}&=&(\rho-\tau)\Phi^{\prime}-\frac{2}{r}(p+\tau),
\label{bphi} \eea which yield, together with (\ref{hawking}) and
(\ref{t0}), the Hawking temperature, wormhole temperature and
wormhole shape function, \bea T_{H}&=&\frac{1}{4\pi
r^{2}}\frac{b(r)-8\pi G
r^{3}\tau(r)}{\left(1-\frac{b(r)}{r}\right)^{1/2}},\nn\\
T_{0}&=&\frac{1}{4\pi r^{2}}\frac{b(r)-8\pi G
r^{3}\tau(r)}{\left(1-\frac{b(r)}{r}\right)^{1/2}}e^{\Phi(r)},
\label{thbtau}\nn\\
b(r)&=&\int_{0}^{r}{\rm d}r^{\pr}~8\pi G r^{\pr 2}\rho(r^{\pr}),
\eea where we have extended the lower bound of $b(r)$ to zero even
though there exists no exotic matter up to the throat $r=r_{0}$.
Note that in the Hawking and wormhole temperatures in
(\ref{thbtau}), one cannot exclude the possibility of the negative
temperature in the case of $b(r)< 8\pi G r^{3}\tau(r)$. Recently,
the negative temperature has been proposed in the de Sitter
geometry~\cite{minic}. However, due to the thermodynamic
instability of the de Sitter space, the negative temperature was
prohibited.  In the wormhole case, there could exist the
``negative temperature" originated from the ``exotic matter,"
different from the de Sitter case which does not allow the exotic
matter.

Next, for the MT wormhole geometry to be connectible to
asymptotically flat space-time, the embedding profile surface
$z=z(r)$ in (\ref{dzdr}) flares outward from the throat at $r=b$
to yield the flaring-out condition at or near the throat \beq
\frac{d^{2}r}{dz^{2}}=\frac{b-b^{\pr}r}{2b^{2}}>0, \eeq from which
one obtains the constraint at or near the throat~\cite{mt88}, \beq
\frac{\tau-\rho}{|\rho|}>0. \label{taurho} \eeq

For the limit of $x^{5}\rightarrow\pm\infty$, one has two regions
(or universes), and at $x^{5}=0$ the wormhole shape function
$b(r)$ has a minimum value at $r=r_{0}$. In order to investigate
the thermodynamics in the wormhole geometry, one may consider a
specific case that the exotic matter resides only on the neck of
the wormhole, \beq \rho (r)=\rho_{0}\delta (r-r_{0}),~~~\tau
(r)=\tau_{0}\delta (r-r_{0}),\label{rhodelta} \eeq and the two
universes contact at $x^{5}=0$. Note that the wormhole shape
function is then given by \beq b(r)=8\pi G r_{0}^{2}\rho_{0}>0,
\eeq to yield the Hawking temperature and wormhole temperature,
\bea T_{H}&=&\frac{b_{0}}{4\pi
r^{2}\left(1-\frac{b_{0}}{r}\right)^{1/2}}-\frac{2Gr_{0}\tau_{0}}
{\left(1-\frac{b_{0}}{r_{0}}\right)^{1/2}}\delta(r-r_{0}), \nn\\
T_{0}&=&\frac{b_{0}}{4\pi r^{2}}-2Gr_{0}\tau_{0}\delta(r-r_{0}),
\label{hawkingdel}\eea showing that the wormhole temperature is
positive (negative) outside (inside) the exotic matter
distribution, since at $r\neq r_{0}$ one has only the positive
first term while at $r=r_{0}$ the negative second term associated
with the delta function and the suppressed first term.  Note that,
even though the metric (\ref{staticmetric}) has smooth geometry,
the temperature (\ref{hawking}) can be discontinuous since the
factor $\Phi^{\pr}(r)$ in (\ref{bphi}) is associated with the
exotic matter to yield discontinuity as in the temperature
(\ref{hawkingdel}).

For the further investigation of the wormhole temperature outside
the exotic matter distribution, we assume for brevity that $\rho$,
$\tau$ and $p$ vanish at all radii $r>r_{0}$ for some surface
radius $r=r_{0}$ with constant distribution for some finite
region. In this distribution $\Phi'$ is also negative which means
``negative temperature'' by the third relation of (\ref{bphi}).
One can thus have outside the cut-off at $r=r_{0}$ the geometry of
the standard Schwarzschild form~\cite{mt88} \bea
b(r)&=&b(r_{0})={\rm const}=B>0,\nonumber\\
\Phi(r)&=&\frac{1}{2}{\rm ln}\left(1-\frac{B}{r}\right).
\label{stand} \eea Exploiting (\ref{hawking}) and (\ref{stand}),
we arrive at the Hawking and wormhole temperatures in the region
$r>r_{0}$ \bea T_{H}&=&\frac{B}{4\pi
r^{2}\left(1-\frac{B}{r}\right)^{1/2}},\nn\\
T_{0}&=&\frac{B}{4\pi r^{2}},\eea which are positive definite
since $\tau$ vanishes in this region. Note that $T_{H}$ and
$T_{0}$ vanish asymptotically consistent with the cosmological
phenomenology.

To figure out further the negative temperature inside the exotic
matter distribution, for the wormhole mass-energy density $\rho$
and tension $\tau$ one may take an ansatz of the
forms~\cite{kim96} \beq
\rho(r)=\rho_{0}r^{\beta},~~~\tau(r)=\tau_{0}r^{\beta},
\label{tbeta} \eeq which, together with the inequality
(\ref{taurho}), yields \beq b(r)<\frac{8\pi G
r^{3}}{\beta+3}\tau(r), \eeq so that the Hawking and wormhole
temperatures (\ref{thbtau}) are negative definite except the case
$-3<\beta<-2$. However, the powers in the interval $-3<\beta<-2$
cannot take place in the tension $\tau(r)$ of the MT wormhole
matter distribution satisfying the asymptotic flaring-out
condition which holds in the interval $\beta<-3$. The Hawking and
wormhole temperatures within the exotic matter thus become
negative at least when one assumes the energy density and the
tension of the form(\ref{tbeta}).

In the wormhole geometry (\ref{rhodelta}), one may have a manifold
of the form $S^{2}\times h$ with $h\in (-\delta,+\delta)$ $(\delta
\rightarrow 0)$ and the wormhole has two-sphere boundaries of
radius $b_{0}$. Moreover, the entropy seen by an accelerated
observer in the Minkowski space is attainable from the transverse
area to the observer~\cite{laf87}. In an embedded higher
dimensional flat manifold, there exist embedding constraints to
yield the finite transverse area or entropy.  In the static
wormhole GEMS of interest, one can thus formulate the entropy
lower bound: \beq S=2\int dx^{2}dx^{3}dx^{4}dx^{5}dx^{6}
[(x^{2})^{2}+(x^{3})^{2}+(x^{4})^{2}-r^{2}] \delta (x^{5})\delta
[x^{6}-f(r)] \label{entropy00} \eeq where $f(r)$ can be read off
from (\ref{staticgems}).  Here the integration over $x^{6}$
subject to the constraint $\delta [x^{6}-f(r)]$ is unity and the
constraint $\delta (x^{5})$ leads to $r=b_{0}$ so that the $x^{5}$
integral yields unity and the remaining integrals over $x^{i}$
($i=2,3,4$) with constraint $\delta
[(x^{2})^{2}+(x^{3})^{2}+(x^{4})^{2}-r^{2}]$ produce the area
$4\pi b_{0}^{2}$.  The factor 2 originates from the fact that one
has two boundaries at $h=\pm \delta$.  We thus arrive at the
entropy lower bound \beq S=8\pi b_{0}^{2}, \label{entropy01} \eeq
which is consistent with the holographic description that all the
microscopic quantum information is deposited on the upper and
lower two-sphere boundaries of radius $b_{0}$, and with the fact
that the entropy is extensive quantity.  Moreover, for the other
geometries of the standard Schwarzschild form (\ref{stand}) and of
the ansatz (\ref{tbeta}), one can have additional area
contributions from the nonvanishing surfaces for the wormhole
shape functions at $r\neq r_{0}$, to yield the increased
entropies. The entropies for any cases are then constrained by the
general ``lower bound" value (\ref{entropy01}).

\noindent {\it 4. Conclusion.} We have studied the (3+1)
Morris-Thorne wormhole to obtain the (5+2) higher dimensional flat
embedding structure.  We have thus shown on these flat embedding
geometries that the wormhole temperature has negative (positive)
values inside (outside) the exotic matter distribution accumulated
mostly around the wormhole shape radius, and the wormhole entropy
lower bound is twice the throat area of the wormhole.  Note that
for the definition of the exotic matter one has the constraint
condition (\ref{taurho}) between $\rho$ and $\tau$, and in this
paper we have assumed several forms of ansatz for these variables
to evaluate the wormhole temperatures inside the exotic matter
distributions. It will be interesting in further
investigation to study the wormhole temperature without any ad hoc ansatz.\\

The authors would like to acknowledge financial support in part
from the Korea Science and Engineering Foundation Grant
R01-2000-00015.

\end{document}